\documentclass[prd,twocolumn,amsmath,aps,floats,amssymb, floatfix, superscriptaddress, nofootinbib]{revtex4}

\pdfoutput=1

\usepackage{graphicx}
\usepackage{wasysym}
\usepackage{amsfonts}
\usepackage{subfigure}
\usepackage{hyperref}

\begin{document}

\title{Inflaton or Curvaton?\\Constraints on Bimodal Primordial Spectra from Mixed Perturbations}
\author{William H. Kinney}
\email[Email:]{whkinney@buffalo.edu}
\affiliation{Department of Physics, University at Buffalo, The State University of New York, Buffalo, NY 14260-1500, USA}
\author{Azadeh Moradinezhad Dizgah}
\email[Email:]{am248@buffalo.edu}
\affiliation{Department of Physics, University at Buffalo, The State University of New York, Buffalo, NY 14260-1500, USA}
\author{A. Riotto} 
\email[Email:]{antonio.riotto@unige.ch}
\affiliation{Department of Theoretical Physics and Center for Theoretical Physics, University of Geneva, \\
 quai E. Ansermet 24, CH-1211 Geneva 4}

\begin{abstract}
\noindent 
We consider Cosmic Microwave Background constraints on inflation models for which the primordial power
spectrum is a mixture of perturbations generated by inflaton fluctuations and fluctuations in a curvaton
field.  If future experiments do not detect isocurvature modes or large non-Gaussianity, it will not be possible to directly distinguish inflaton and curvaton
contributions.  We investigate whether current and future data can instead constrain the relative
contributions of the two sources. 
We model the spectrum with a bimodal form consisting of a sum of two independent power laws, with
different spectral indices. We quantify the ability of current and upcoming data sets to constrain the
difference $\Delta n$ in spectral indices, and relative fraction $f$ of the subdominant power spectrum at a
pivot scale of $k_0 = 0.017\ {{\rm Mpc^{-1}} \ h}$. Data sets selected are the WMAP 7-year data, alone and in conjunction with South Pole Telescope data, and a synthetic data set comparable to the upcoming Planck data set. We find that current data show no increase in quality of fit for a mixed inflaton/curvaton power spectrum, and a pure power-law spectrum is favored. The ability to constrain independent parameters such as the tensor/scalar ratio is not substantially affected by the additional parameters in the fit. Planck will be capable of placing significant constraints on the parameter space for a bimodal spectrum. 
\end{abstract}

\maketitle

\section{Introduction}
\noindent
Current astrophysical data are consistent with a primordial spectrum of cosmological density perturbations
consisting of a power-law spectrum, with spectral index close to scale invariance. This is consistent with
the predictions of the simplest single-field models of cosmological inflation. In the single-field
scenario, fluctuations in the field responsible for inflation (the {\it inflaton}) generate a spectrum of
curvature perturbations in the early universe. In this case, the spectral index of the resulting
perturbation spectrum is determined by the shape of the inflaton potential. However, other mechanisms for
generating curvature perturbations are possible, for example the {\it curvaton} scenario
\cite{Linde:1996gt,Enqvist:2001zp,Lyth:2001nq,Moroi:2001ct}, in which a massless ``spectator'' field sources isocurvature perturbations during the period of inflationary expansion. The isocurvature perturbations are subsequently converted to curvature perturbations when the curvaton field decays, which is assumed to happen after the end of inflation and reheating. Like single-field inflation, the simplest curvaton models predict a power-law spectrum of perturbations, but in the curvaton case, the shape of the spectrum is completely unrelated to the inflaton potential. 

Even when a curvaton is present, however, inflaton fluctuations will still generate curvature perturbations
\cite{Langlois:2004nn}. While it is typically assumed that perturbations from the inflaton are subdominant to the curvaton spectrum, there is no physics requiring this to be so. It is in principle possible for both power spectra to be of similar amplitude -- in fact, suppressing the inflaton spectrum requires fine-tuning of the inflaton self coupling even beyond the $\lambda \sim 10^{-14}$ already required by Cosmic Microwave Background (CMB) normalization, so it is reasonable to expect that inflaton-generated perturbations will be significant even when a curvaton is present. 
An important distinction between inflaton and curvaton-sourced perturbations is that, while the curvaton three-point correlator, the so-called bispectrum, can be sizable \cite{Lyth:2002my,Bartolo:2003jx}
, that of the inflation is suppressed by slow-roll parameters and is therefore totally negligible \cite{Acquaviva:2002ud,Maldacena:2002vr}. A future detection of
non-Gaussianity (NG) in the CMB will therefore point to a light field like the curvaton as the source of the curvature perturbation \cite{Bartolo:2004if}. Additionally, a residual isocurvature component in the primordial spectrum will indicate the presence of the curvaton \cite{Easson:2010uw}.  However, suppose that future experiments do not give any sign of NG or isocurvature modes. How can we know if the curvature perturbations come from the inflaton or if they are 
created by a different mechanism, {\it e.g.} by converting isocurvature perturbations into adiabatic ones? This is an empirically relevant question, since single field models of inflation will be constrained under the assumption that the curvature perturbations originated
from the inflaton. The result is that a given set of CMB observables can be mapped to a wide range of different inflaton
potentials, each with a different, unresolved curvaton contribution.  This inversion was studied in Ref.
\cite{Easson:2010uw}, where it was found that the resulting degeneracy in the inflaton potential parameter
space is large.  While additional observables beyond the adiabatic and tensor two-point correlators are
ultimately needed to disentangle the inflaton and curvaton contributions, in this work we examine whether
current and future data can instead shed light on their relative contributions to the overall curvature
spectrum.

In this paper we therefore consider CMB constraints from current and upcoming data on a mixed inflaton/curvaton primordial power spectrum, which we model as a superposition of two uncorrelated power laws with independent spectral indices. The paper is structured a follows: Sec. \ref{sec:Model} specifies the model studied and defines the parameters fit to CMB data. Section \ref{sec:Constraints} presents constraints from the WMAP 7-year (WMAP7) data, WMAP7 plus South Pole Telescope (SPT) data, and simulated Planck-quality data consistent with current constraints. Section \ref{sec:Conclusions} summarizes our conclusions. 

\section{Bimodal Power Spectrum}
\label{sec:Model}
\noindent
In this work we consider scalar perturbations formed from the superposition of two
uncorrelated power law spectra, 
\begin{equation}
\label{def}
P(k) = A_1 \left(\frac{k}{k_0}\right)^{n_1 -1} +  A_2 \left(\frac{k}{k_0}\right)^{n_2-1}.
\end{equation}
Bimodal spectra of this form can arise if both the inflaton and an additional decoupled
field contribute to the overall curvature spectrum.  As a specific physical model to
motivate Eq. (\ref{def}), we consider mixed
inflaton and curvaton fluctuations.  The curvaton, $\sigma$, is a weakly coupled scalar field that is relatively light during inflation ($m^2_\sigma \ll H^2$).  It is decoupled from the inflationary dynamics and over-damped by the expansion,
$3H\dot{\sigma}\simeq 0$, until $m_\sigma^2 \lesssim H^2$.  After inflation, the curvaton is free to
roll to the minimum of its potential, $V(\sigma)$, where it undergoes oscillations during a radiation
dominated phase.  
The fluctuations in the energy densities of the radiation and curvaton field generate curvature perturbations,
$\psi_i$, given in terms of the gauge invariant expression,
\begin{equation}
\zeta_i = -\psi - H\frac{\delta \rho_i}{\dot{\rho}_i}.
\end{equation}
The variable $\zeta_i$ reduces to the curvature perturbation on uniform density hypersurfaces.
When the curvaton decays into radiation, the total adiabatic perturbation is 
\begin{eqnarray}
\zeta &=& r_\sigma \zeta_\sigma + (1-r_\sigma)\zeta_{\rm inf}  \nonumber \\
      &=& \zeta_{inf} + \frac{r_\sigma}{3} S_\sigma
\end{eqnarray}
where ``$\sigma$'' and ``inf'' denote the curvaton and inflaton perturbations, respectively, and 
\begin{equation}
r_\sigma = \frac{3\Omega_\sigma}{4- \Omega_\sigma}
\end{equation}
gives the relative contribution of the curvaton to the total curvature perturbation at the time of decay. $S_\sigma$ is the non-adiabatic part of curvaton perturbations and is defined as $S_\sigma \equiv 3(\zeta_\sigma-\zeta_{inf})$. From this expression it is evident that
the contribution of the curvaton grows as it oscillates in the post-inflationary universe before decay: an oscillating scalar evolves as
$\rho_\sigma \propto a^{-3}$, and radiation as $\rho_\gamma \propto a^{-4}$, giving $\rho_\sigma/\rho_\gamma \approx
\Omega_\sigma \propto a$. 

Since the adiabatic inflaton perturbations and the isocurvature curvaton fluctuations are uncorrelated, the total power spectrum of
primordial curvature perturbations is given by
\begin{eqnarray}
\label{fullspec}
P(k) &=& \frac{k^3|\zeta|^2}{2\pi^2} \nonumber\\
&=& P_{\rm inf}(k) + \frac{r_\sigma^2}{9} {P_S}_{\sigma}(k).
\end{eqnarray}

One can therefore model the total power spectrum of the mixed curvaton-inflaton perturbations to lowest order in slow-roll, as a combination of two power laws
\begin{equation}
\label{def1}
P(k) = A_s\left[(1-f)\left(\frac{k}{k_0}\right)^{n_{\rm inf}-1}+
f\left(\frac{k}{k_0}\right)^{n_{\sigma }-1}\right],
\end{equation}
which is of the form Eq. (\ref{def}) with 
\begin{eqnarray}
A_s &=& \frac{P_{\rm inf}(k_0)}{1-f}, \\
f&=& \frac{r_\sigma ^2 {P_S}_\sigma(k_0)}{9 P_{inf}(k_0) + r_\sigma^2 {P_S}_\sigma(k_0)}. 
\end{eqnarray}
The function $f$ has been introduced to control the relative contributions of the two
spectra, allowing the composite spectrum to range from pure curvaton ($f=1$) to pure inflaton ($f=0$). 
In the absence of additional discriminating observables like isocurvature modes or
NG, it is generally not possible to separately constrain the curvaton and inflaton
contributions; the resulting degeneracy is not easily reduced \cite{Easson:2010uw}. We emphasize that the assumption of a pure power law for both the inflaton and curvaton components is restrictive: in principle, either spectrum could also have intrinsic scale dependence, which would add additional parameters to the fit.

The effective spectral index and running of the full bimodal spectrum Eq. (\ref{def1}) can be obtained
by Taylor expanding the spectrum in $\ln k$.  From the definitions
\begin{eqnarray}
\frac{{\rm d} \ln P(k)}{{\rm d} \ln k} &\equiv& n_s - 1,\\
\frac{{\rm d^2} \ln P(k)}{{\rm d} \ln k^2} &\equiv& \alpha,
\end{eqnarray}
we obtain the relations,
\begin{eqnarray}
\label{index}
n_s - 1 &=& n_{\rm inf}-1 - \frac{\alpha}{1-f},\\
\label{run}
\alpha &=& (1-f)f(\Delta n)^2,
\end{eqnarray}
where $\Delta n = n_{\rm inf}-n_\sigma$. 
\begin{figure}[htp]
\includegraphics[width=0.49 \textwidth,clip]{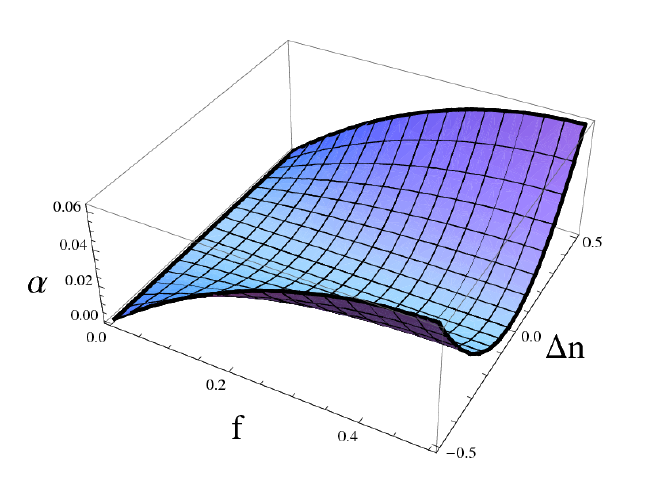}
\caption{Depiction of $\alpha$ as a function of $\Delta n$ and $f$.  Running is largest when both $\Delta n$ and
$f$ are large.}     
\label{fig:alpha}
\end{figure} 
One notable feature of the spectrum is that the
effective running is positive, $\alpha > 0$, a simple consequence of the fact that the component with lower spectral index will always dominate on large enough scale, and the component with higher spectral index will always dominate on small scales. This is in tension with current cosmic microwave
background measurements which indicate a preference for $\alpha < 0$
\cite{Komatsu:2010fb,Dunkley:2010ge,Keisler:2011aw}.  This fact can be used to constrain the
bimodal spectrum, as a model, relative to the power law spectrum of the pure curvaton or inflaton.
And, while measurements of $n_s$ and $\alpha$ are not sufficient (by themselves) to determine the underlying
parameters $n_{\rm inf}$, $n_\sigma$, and $f$, Eqs. (\ref{index}) and
(\ref{run}) can be used with precision measurements of $n_s$ and $\alpha$ to establish
useful constraint relations between the underlying parameters.  Additional observables, like isocurvature modes or NG,
can be used to further improve the reconstruction of the bimodal spectrum if they are observed in the future.  We
consider this possibility as well. 

\section{Constraints from CMB Measurements}
\label{sec:Constraints}
\noindent
In this section we obtain constraints on the bimodal power spectrum Eq. (\ref{def1}).  It is assumed that $n_{\rm inf}$ and $n_\sigma$ share the same prior range so that $f$ need only vary between 0 and 0.5.  This restriction gives the parameterization  
 \begin{equation}
\label{def2}
P_\mathcal{R}(k) = A_{s} \left[(1-f)\left(\frac{k}{k_0}\right)^{n_{\rm dom} -1} +  f \left(\frac{k}{k_0}\right)^{n_{\rm sub}-1}\right],
\end{equation}
where ``dom'' labels the dominant and ``sub'' the sub-dominant spectrum, measured at a pivot scale of $k_0 = 0.017\ {{\rm Mpc^{-1}} \ h} $. 
We use the 7-year WMAP dataset (WMAP7) in combination with data from the South Pole Telescope (SPT) \cite{Komatsu:2010fb,Keisler:2011aw} to perform a Bayesian analysis of the power spectrum in terms of the underlying
parameters $A_{s}$, $f$, $n_{\rm dom}$, and $n_{\rm sub}$.  The base cosmological parameters are
chosen to be the baryon and cold dark matter densities, $\Omega_b h^2$ and $\Omega_c h^2$, the
ratio of the sound horizon to the angular diameter distance at decoupling, $\theta_s$, and the
optical depth to reionization, $\tau$.  We also vary the tensor/scalar ration, $r$.  Use of the SPT data introduces three more foreground
parameters: the Poisson point source power from randomly distributed galaxies, $D^{\rm PS}_\ell$,
the clustered point source power, $D^{\rm CL}_\ell$, and the secondary emission from the
Sunyaev-Zeldovich effect from clusters, $A_{\rm SZ}$. We also obtain constraints for a simulated
Planck-precision data set that we describe below.  We perform the parameter estimation using Markov Chain Monte Carlo
(MCMC) with \texttt{CosmoMC} \cite{Lewis:2002ah}, marginalizing over the foreground parameters.
Chain convergence was determined with the Gellman-Rubin statistic, $R-1<0.1$. 

While ground-based observatories like the Atacama Cosmology Telescope (ACT) and SPT have provided increasingly precise
measurements of $\alpha$, ESA's Planck Surveyor satellite promises the best
constraints yet on both $n_s$ and $\alpha$.  We therefore obtain constraint projections
on the bimodal spectrum based on a simulated Planck-precision data set in addition to WMAP7+SPT.
We perform MCMC on the same set of base cosmological parameters as above:  $\Omega_b
h^2$, $\Omega_c h^2$, $\theta_s$, $\tau$, with spectral parameters $A_s$, $n_{\rm dom}$,
$n_{\rm sub}$, and $f$, using simulated temperature and
polarization data out to $\ell = 2000$. We conduct the analysis on a fiducial model with a power law spectrum and
$r=0$.
The base cosmological parameters ($\Omega_b h^2$, $\Omega_c
h^2$, $\theta_s$, $\tau$, $A_s$, and $n_s$) are taken equal to
the WMAP7+SPT best-fit values.   
Data is simulated for three channels with frequencies (100 GHz, 143 GHz, 217 GHz) and noise levels per Gaussian beam
$(\sigma^{T}_{\rm pix})^2 =$ ($46.25$ $\mu$K$^2$, $36$ $\mu$K$^2$, $17.6$ $\mu$K$^2$) (with $\sigma_{\rm pix}^P =
\sqrt{2}\sigma_{\rm pix}^T$ for polarization spectra).  The FWHM of the three channels are $\theta_{\rm fwhm} =$ (9.5',
7.1', 5.0') and we assume a sky coverage fraction of 0.65 \cite{planck:2006uk}. 
\begin{figure*}
\centering
$\begin{array}{ccc}
\subfigure[]{
\includegraphics[width=0.5 \textwidth,clip]{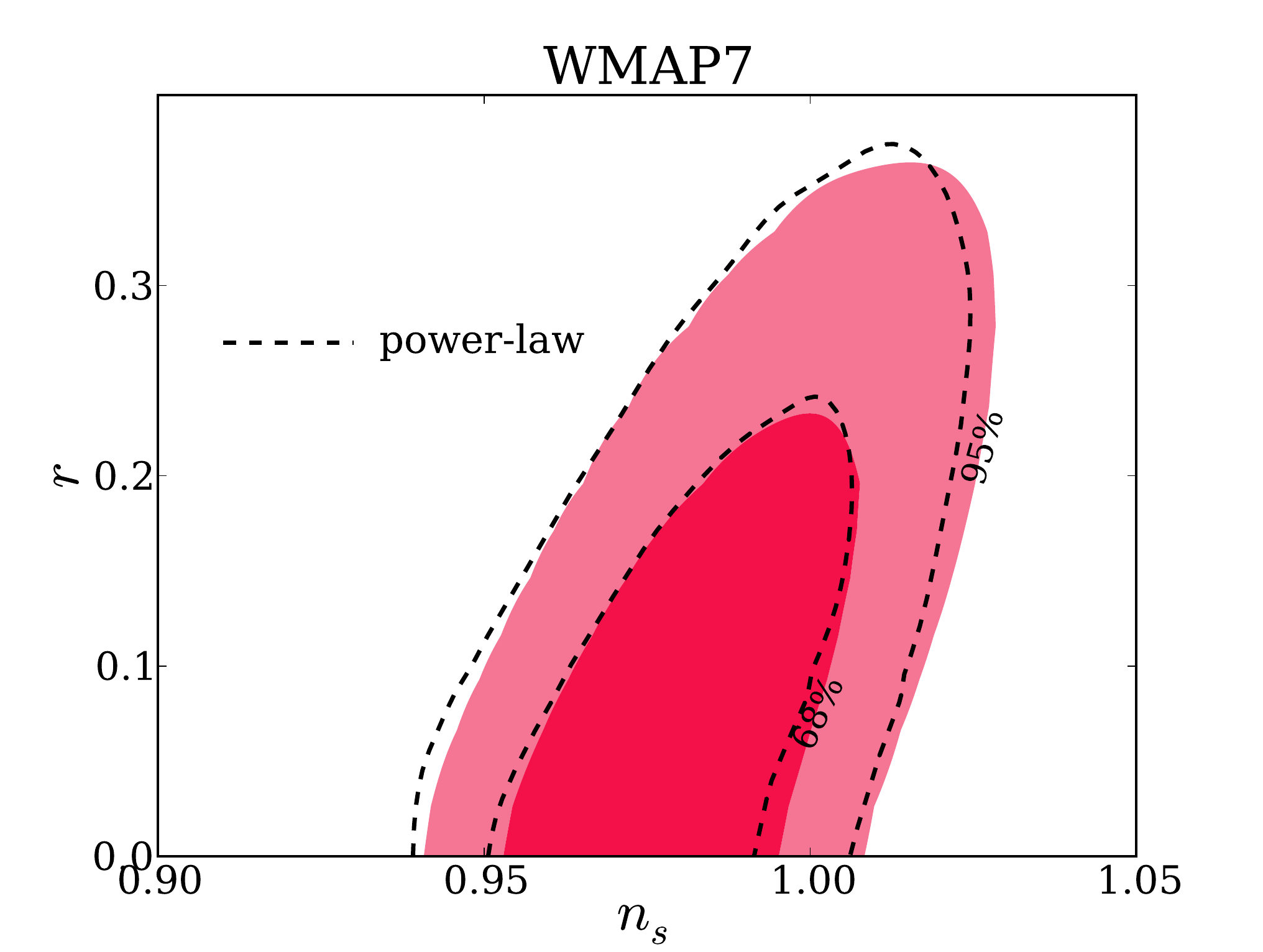}}
\subfigure[]{
\includegraphics[width=0.5 \textwidth,clip]{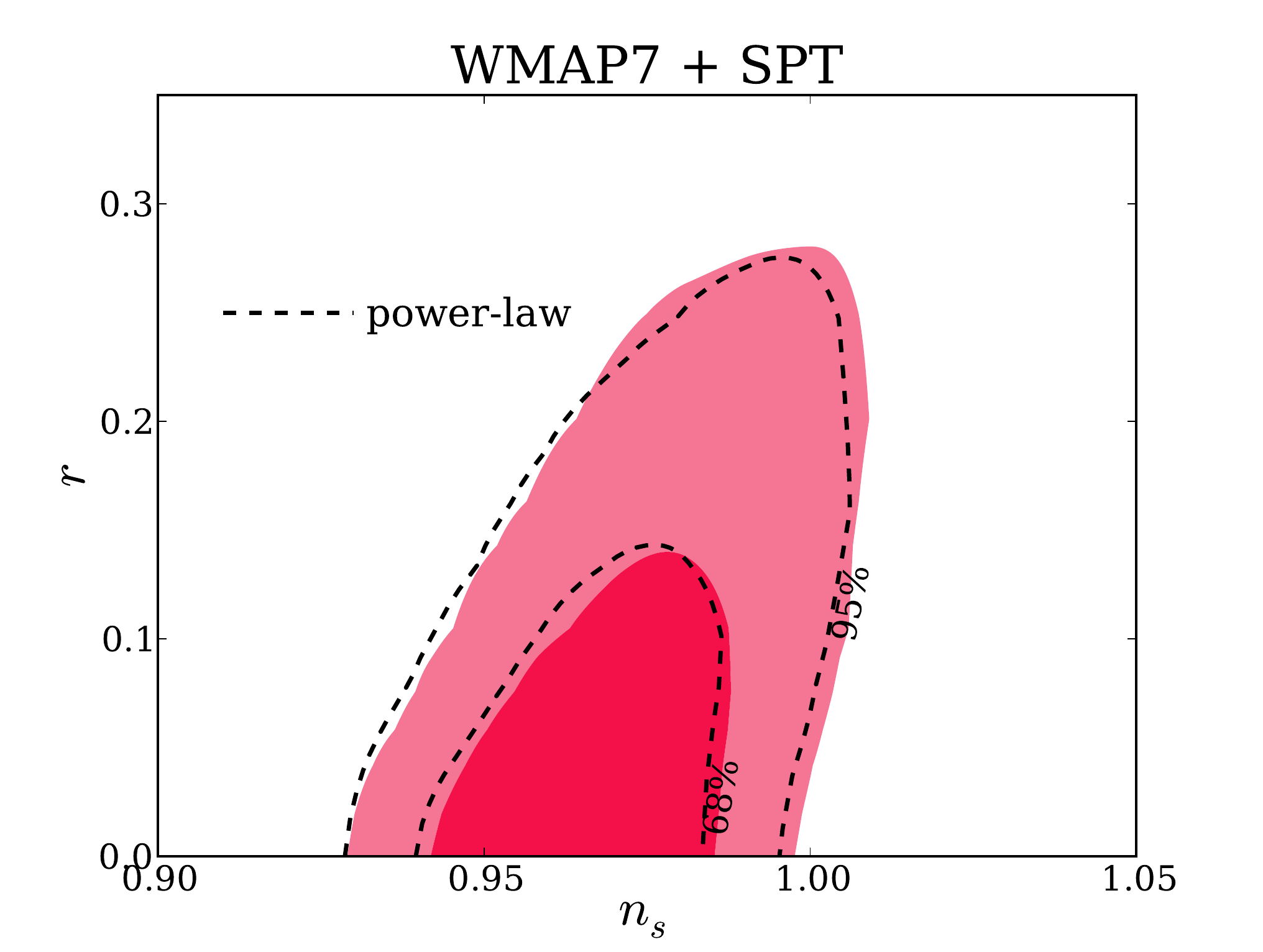}}
\end{array}$
\caption{Marginalized constraints in the effective $n_s$-$r$ plane for the bimodal spectrum for (a) WMAP7 and (b) WMAP7+SPT.  For each data set we also include results from a power law fit (dashed lines).  The two sets of
contours overlap in each case.}     
\label{fig:nr}
\end{figure*} 
\begin{figure}
\includegraphics[width=0.5 \textwidth,clip]{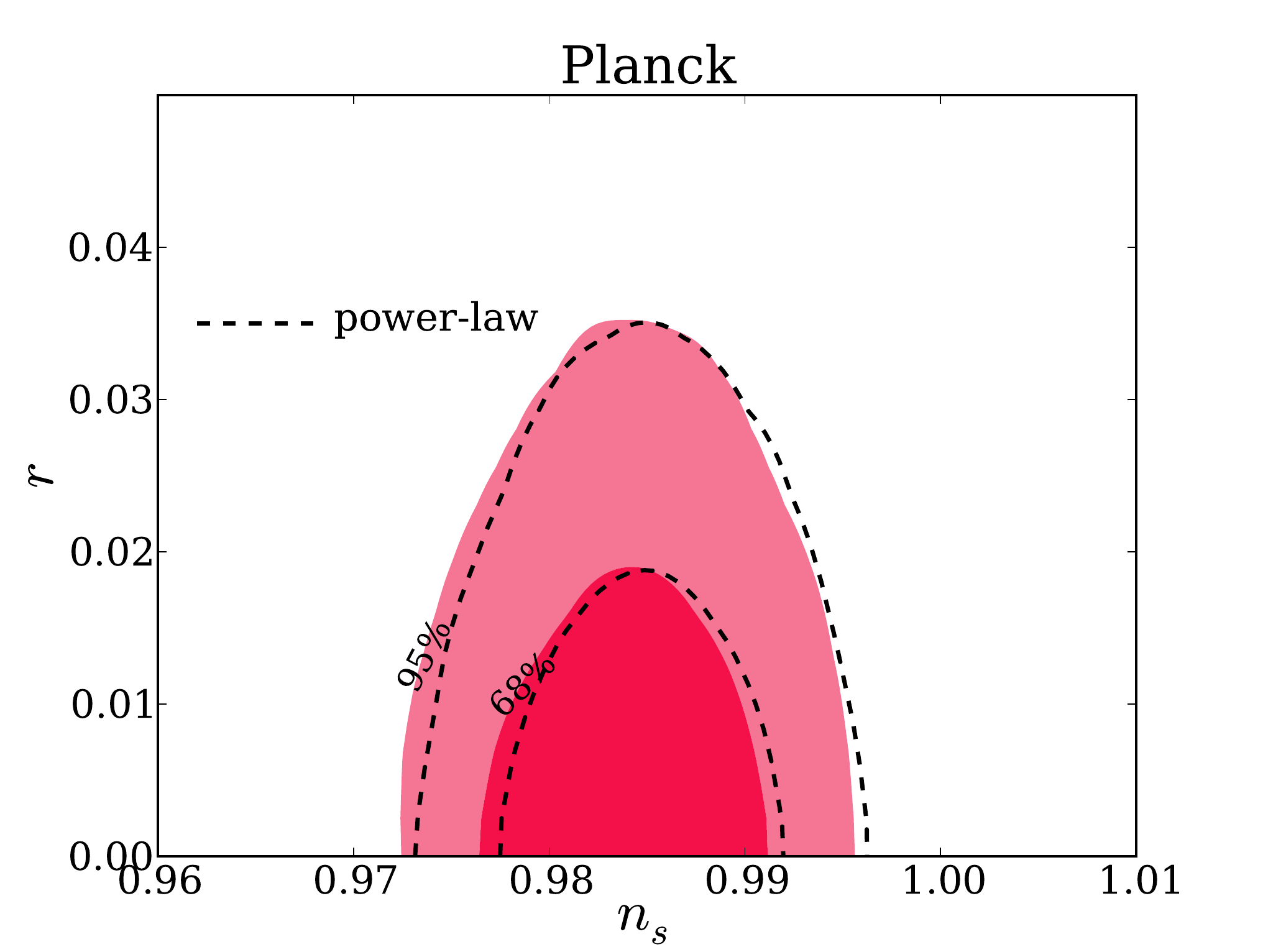}
\caption{Marginalized constraints in the effective $n_s$-$r$ plane for the bimodal spectrum for a simulated Planck-precision data set, showing a power-law fit (dashed lines).}
\label{fig:nrPlanck}
\end{figure}

Although the MCMC furnishes constraints directly on the base parameters of Eq. (\ref{def2}), we first examine
estimates for the derived parameter $n_s$.  We compare constraints in the $n_s$-$r$ plane
obtained for a power law ansatz, $P(k) \propto k^{n_s - 1}$, with those obtained for the bimodal spectrum Eq.
(\ref{def2}).  The bimodal spectrum can be written as a single spectrum with an effective spectral index and
running, obtained from the base parameters via Eq. (\ref{index}).  The bimodal spectrum therefore effectively has
one more free parameter than the power law, and in general, the addition of a free parameter can improve a model's fit to data simply on
account of the reduced number of degrees of freedom.  If the additional parameter has correlations with the
base parameters, degeneracies might open up and constraints on the correlated parameters might
degrade.  This is what happens in the $n_s$-$r$ plane when running is added as a free parameter.  
We present results in Figures \ref{fig:nr} and \ref{fig:nrPlanck} for the three data sets: (a) WMAP7, (b) WMAP7+SPT, and (c) Planck.  Interestingly, 
none of the constraints obtained here suffer from any such degradation, with the contours in each case virtually
overlapping.  This is because the additional effective
parameter of the bimodal spectrum is {\it positive} running, which is not well-tolerated by current CMB data:
-0.060 $<$ $\alpha$ $<$ -0.008 (WMAP) \cite{Komatsu:2010fb} and -0.037 $<$ $\alpha$ $<$ -0.011 (WMAP+SPT)
\cite{Keisler:2011aw} at 68\% CL, both assuming a zero tensor amplitude. Constraints are weaker when $r$ is included as a joint parameter in the fit \cite{Kinney:2008wy}. For the bimodal model, the points of appreciable likelihood -- those that
contribute to the contour -- are those for which $\alpha \approx 0$.  At the pivot scale ($k_0 = 0.017\,h{\rm Mpc}^{-1}$), positive running forces a red spectral index.
A red spectral index with positive running will fit the data worse than a red spectral index with no running,
because the positive running results in an excess of power at the low multipoles and a blue spectrum at high
multipoles.  These features are not in agreement with CMB data \cite{Hlozek:2011pc}.

This fact is also revealed by examining the overall model fits: 
the bimodal spectrum gives a best-fit log-likelihood $-2\ln \mathcal{L} =
\chi^2_{eff}=7514.25$, while the power law gives $\chi^2_{eff} = 7514.4$ for WMAP7+SPT, with similar close agreement for the other data sets.  The reduced $\chi^2_{eff}$,
given by $\chi^2_{eff}/\nu$, is the relevant statistic for model comparison.   Here $\nu$ denotes the number of
degrees of freedom, defined as the number of data points minus the number of free parameters. With two fewer
degrees of freedom ({\it i.e.} two additional parameters), the bimodal spectrum gives a larger reduced $\chi^2_{eff}$ and is consequently a worse fit to the data.  
{\it Power law spectra, whether due to a pure inflaton or curvaton, are
therefore preferred over the bimodal spectrum arising from mixed perturbations.}
\begin{figure*}
\centering
$\begin{array}{ccc}
\subfigure[]{
\includegraphics[width=0.32 \textwidth,clip]{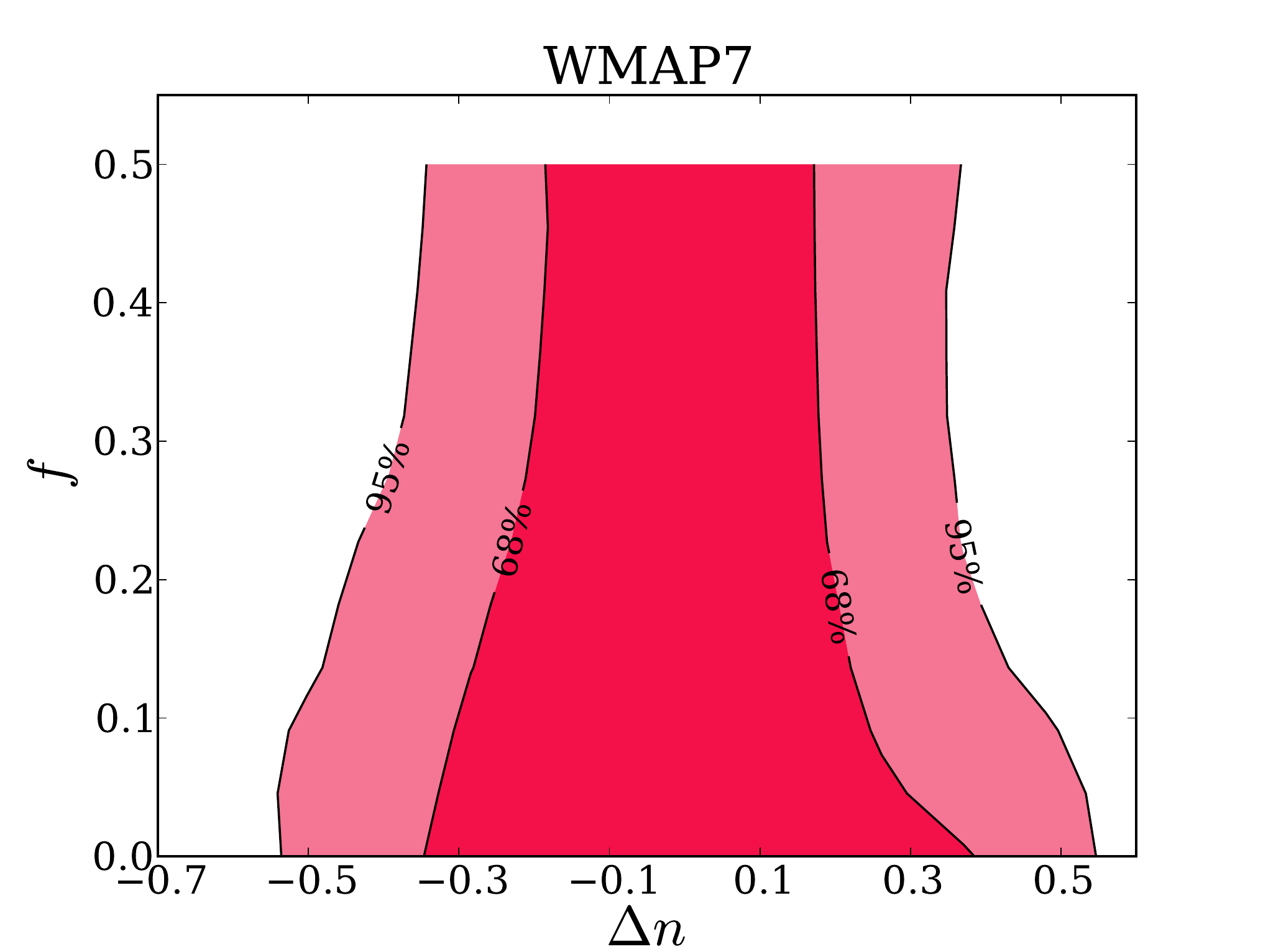}}
\subfigure[]{
\includegraphics[width=0.32 \textwidth,clip]{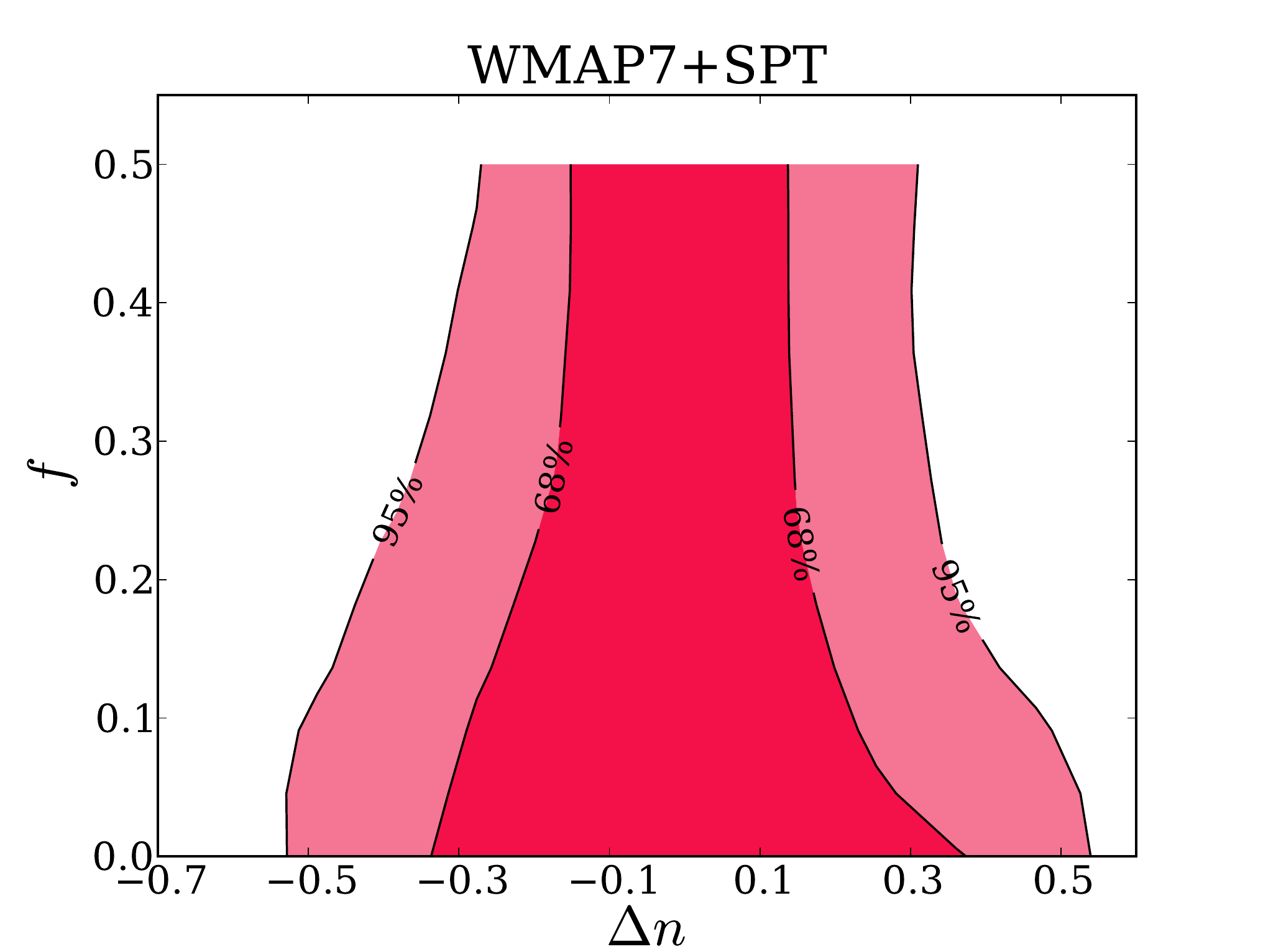}}
\subfigure[]{
\includegraphics[width=0.32 \textwidth,clip]{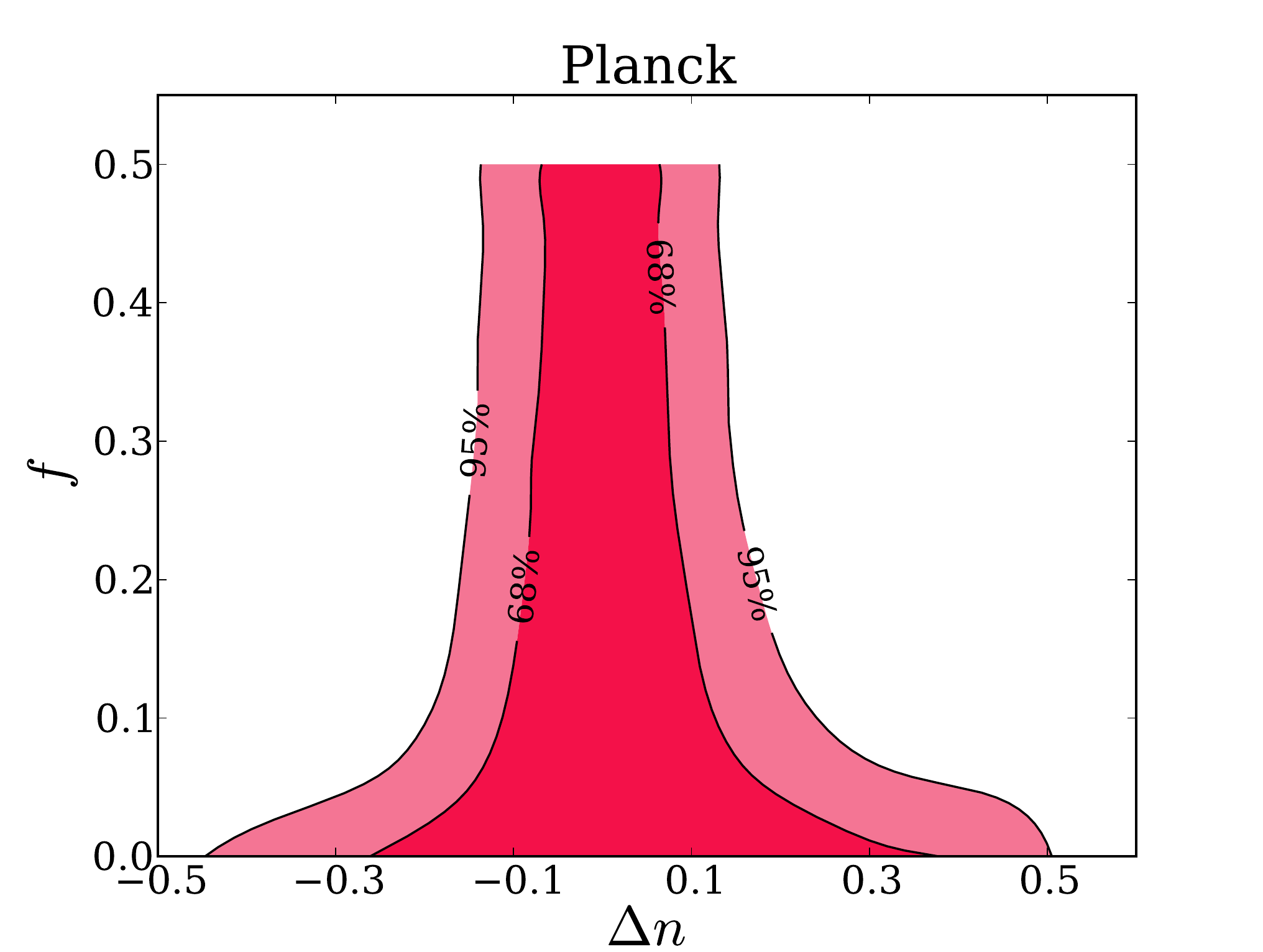}}
\end{array}$
\caption{Marginalized constraints in the $\Delta n$-$f$ plane for (a) WMAP7, (b) WMAP7+SPT, and (c) projection for
a Planck-precision CMB data set.}     
\label{fig:fdeltan}
\end{figure*}

Goodness-of-fit aside, current data does not go so far as to rule out the bimodal power
spectrum.  It is therefore of interest to examine how well CMB measurements constrain
the parameters describing the bimodal power spectrum, since these quantities
are directly related to the underlying inflationary degrees of freedom.  As mentioned,
the base parameters $n_{\rm dom}$, $n_{\rm sub}$, and $f$ cannot be separately
determined from the overall power spectrum, but quality measurements of $n_s$ and $\alpha$ can establish constraint relations
among them.  From Eq. (\ref{run}), a quality  measurement
of running will furnish a relation connecting $\Delta n = n_{\rm dom} - n_{\rm sub}$ and
the fraction $f$.  
We present results in the $\Delta n$-$f$ planes in
Figures \ref{fig:fdeltan} (a), (b), and (c).   
The distributions are shaped by the quality of the constraint on positive running: the inclusion of SPT data
improves the constraint relative to WMAP7 alone,\footnote{It is notable that the improvement in constraints in Figure \ref{fig:fdeltan} (b)
relative to (a) is less striking than would be naively expected given the rather substantial improvement in
constraints on $\alpha$ from SPT.  The effect is one of priors: constraining the bimodal spectrum is
equivalent to fitting a power-law with positive running; the prior on $\alpha$ corresponds to a region of
low relative likelihood where $\alpha$ is not sharply constrained} with Planck offering the highest precision.  With Planck, it becomes possible to
obtain constraints on the fraction $f$: for very different spectral indices $n_{\rm dom}$ and
$n_{\rm sub}$, the fraction is constrained to be very small.  Conversely, for small $f$, the spectrum determined by
$n_{\rm dom}$ completely dominates and $n_{\rm sub}$ is unconstrained.  For $f=0.5$, constraints are symmetric
about $\Delta n = 0$.  Meanwhile, for intermediate $f$, $\Delta n<0$ means that the dominant spectrum is redder at the pivot,
with the subdominant spectrum coming to dominate on smaller scales, leading to a milder overall positive
running.  For intermediate $f$ and $\Delta n>0$, the dominant spectrum is bluer at the pivot than the subdominant
spectrum, which might therefore come to dominate on large scales, again leading to a mild overall positive running.  These two
cases should not, however, be equally favored by the data owing to the fact that SPT places strong constraints
on the high-$\ell$ spectrum, while the low-$\ell$'s remain poorly constrained due to cosmic variance.  There
appears to be a slight indication of this asymmetry in our results, most pronounced in Figure \ref{fig:fdeltan} (c) where constraints on
$\Delta n$ are not symmetric about zero, and tighter on $\Delta n < 0$.

While current data provide no evidence for mixed perturbations, and while future data sets will begin to resolve the bimodal spectrum, the inflaton/curvaton degeneracy remains unbroken.
It is well known that, in general, additional observables beyond the adiabatic and tensor power spectra on CMB
scales are needed to resolve the degeneracy.  It is shown in \cite{Easson:2010uw} that a
detection of isocurvature perturbations would not only break the degeneracy but would give the correct number of
observables to successfully resolve the parameters of the underlying model.  In particular, a measurement of the
correlation angle between the adiabatic and isocurvature components, 
\begin{equation}
\beta \equiv {\rm cos}\Delta =\frac{P_{\zeta \mathcal{S}}(k_0)}{\sqrt{P_\zeta(k_0)P_\mathcal{S}(k_0)}},
\end{equation}
will provide sufficient information to determine the fraction $f$.  The curvaton can generate an isocurvature perturbation in a variety of ways, and the degree of correlation between the isocurvature and adiabatic modes
depends on the process. For
example, a 
cold dark matter isocurvature mode can be set up if the dark matter is a direct product of curvaton decay. In the
pure curvaton limit, this mode is anticorrelated ($\beta = -1$) with the adiabatic mode.  The correlation angle can be written \cite{Kawasaki:2008pa}
\begin{equation}
\beta = -\frac{1}{\sqrt{1+\lambda^{-1}}},
\end{equation}
where $\lambda$ parameterizes the contribution of the curvaton to the overall spectrum, $P_\zeta =
(1+\lambda)P_{\rm inf}$.  The parameter $\lambda$ is related to the fraction $f$ through
\begin{eqnarray}
f &=& \frac{\lambda}{\lambda + 1},\nonumber \\
 &=& \beta^2.
\end{eqnarray}
The degree of correlation varies with the fractional contribution of the curvaton to the overall spectrum: for
anticorrelated modes, $f=1$ as expected.  While current CMB data shows no evidence for an isocurvature mode, it is
still possible that isocurvature perturbations represent a small fraction of the overall density perturbation.  If an
isocurvature mode is detected by Planck, the correlation angle is expected to be measured with a precision of $\delta
\beta \sim 0.04$ at 68\% CL. This leads to a constraint of $\delta f = 2\beta \delta \beta$ on the curvaton
fraction, which, from Figure 3 (c) for $f \gtrsim 0.1$, gives $|\Delta n| \approx 0.1$ at 68\% CL. 
 
Another possibility of resolving the inflation/curvaton degeneracy is to detect a significant 
level of NG. If we parametrize the NG by the dimensionless parameter $f_{\rm NL}\sim \langle\zeta^3\rangle/\langle\zeta^2\rangle^2$ and assume that NG is sourced only by the curvaton, we find 
 \begin{equation}
f_{\rm NL}\sim \frac{f^2}{r_\sigma}.
 \end{equation}
The Planck experiment is sensitive to values of $f_{\rm NL}$ as small as ${\cal O}(5)$. In the case in which no NG is measured,  we conclude that $ r_\sigma\gtrsim f^2/5$.
 A possible detection of 
a contribution to the power spectrum from a second contribution at the
 10\% level would then require $r_\sigma \gtrsim 0.002$.

\section{Conclusions}
\label{sec:Conclusions}
\noindent
In this paper, we consider CMB constraints on a mixed inflaton/curvaton power spectrum, which we model as a
superposition of two power laws with independent spectral indices. Locally, such a bimodal spectrum can be
well approximated by scale-dependent power spectrum with positive running $\alpha$, since the redder
(smallest $n$) spectrum will dominate at large scale, and the the bluer (largest $n$) spectrum will
dominate at small scale (Fig. \ref{fig:alpha}). This is significant in the context of current data, which
favor either a pure power law or negative running \cite{Komatsu:2010fb,Dunkley:2010ge,Keisler:2011aw}.
Consistent with this expectation, we find that adding new parameters describing an effective positive
running does nothing to improve the quality of the fit. 
In particular, constraints on the effective spectral index $n_s$ vs. tensor/scalar ratio, $r$, are almost completely unaffected by the addition of the new parameters (Figs. \ref{fig:nr}, \ref{fig:nrPlanck}), consistent with a single power-law spectrum being favored by the data. In principle, either the inflaton or the curvaton spectrum could have an intrinsic scale dependence, which would complicate the analysis relative to simple assumption of two pure power laws assumed here. This would further reduce the discriminating power of the data.

The parameters $f$, specifying the relative amplitudes of the power spectra, and the difference in spectral
indices $\Delta n$, are poorly constrained (Fig. \ref{fig:fdeltan}). This is true both for the WMAP7 data
considered alone and in combination with South Pole Telescope. 
In the case of the simulated Planck data, we find that Planck will be able to significantly constrain the fraction $f$ in a subdominant spectrum as long as the spectral indices differ by a sufficient amount, $\left\vert\Delta n\right\vert > 0.2$. In this case, a contribution to the power spectrum from a second power law can be ruled out at better than the 10\% level at 95\% confidence by Planck (Fig. \ref{fig:fdeltan}). For $\left\vert\Delta n\right\vert < 0.2$, constraints from Planck will be poor, which is to be expected, since in the limit that both power spectra are identical ($\left\vert\Delta n\right\vert \ll 1$) the spectrum is effectively no longer bimodal, and no constraint on $f$ is possible. 
\newline
\newline

\section*{Acknowledgements}
This research is supported  in part by the National Science Foundation under grants NSF-PHY-0757693 and NSF-PHY-1066278. WHK thanks the Kavli Institute for Cosmological Physics, where part of this work was completed, for generous hospitality.

\end{document}